\documentclass[onecolumn,showkeys,amsmath,amssymb,amsfonts,a4paper,aps,prd,10pt]{revtex4}
\usepackage[colorlinks, linkcolor={blue},citecolor={red}]{hyperref}
\usepackage{graphicx}
\usepackage{epstopdf}
\newcommand{\nc}{\newcommand}
\nc{\ba}{\begin{eqnarray}} \nc{\ea}{\end{eqnarray}}
\newcommand\be{\begin{equation}}
\newcommand\ee{\end{equation}}
\nc{\D}{\overline{\mbox{D3}}}

\nc{\ga}{\gamma} \nc{\tnu}{\tilde{\nu}} \nc{\tmu}{\tilde{\mu}}

\nc{\x}{{\bf{x}}}

\input{epsf.sty}
\begin{document}
\title{Metric Measure Space as a Framework for Gravitation}
\author{Nafiseh Rahmanpour}
\email{n$_$rahmanpour@sbu.ac.ir}
\author{Hossein Shojaie}
\email{h-shojaie@sbu.ac.ir}
\affiliation{Department of Physics, Shahid Beheshti University, G.C., Evin, Tehran 1983969411, Iran}

\begin{abstract}

In this manuscript, we show how conformal invariance can be incorporated in a classical theory of gravitation, in the context of metric measure space. Metric measure space involves a geometrical scalar $f$, dubbed as density function, which here appears as a conformal degree of freedom. In this framework, we present conformally invariant field equations, the relevant identities and geodesic equations. In metric measure space, the volume element and accordingly the operators with integral based definitions are modified. For instance, the divergence operator in this space differs from the Riemannian one. As a result, a gravitational theory formulated in this space has a generalized second Bianchi identity and a generalized conservation of energy-momentum tensor. It is shown how, by using the generalized identity for conservation of energy-momentum tensor, one can obtain a conformally invariant geodesic equation. By comparison of the geodesic equations in metric measure space with the Bohmian trajectories, in both relativistic and non-relativistic regimes, a relation between density function $f$ and the quantum potential is proposed. This suggests metric measure space to be considered as a suitable framework for geometric description of Bohm's quantum mechanics. On the other hand, as it is known, Weyl geometry is one of the main approaches to construct conformally invariant gravitational models. Regarding the fact that the connection in the integrable Weyl space is modified and in metric measure space remains the same as it is in the Riemann space, the mathematical analogy between these two spaces is also discussed. 
\end{abstract}
\maketitle


\section{Introduction}

Metric measure space is a generalization of Riemann space wherein the measure (volume element) is considered as $e^{-f}dvol(g)$. In this space, in addition to the metric $g$ a geometrical scalar field $f$, which is called density function, is employed. Metric measure space, in the mathematical literature, is used in many different studies, from the collapsing manifolds with Ricci curvature bounded below to Perelman's approach in the realization of Ricci flow as a gradient flow~\cite{Cheeger, Cheeger2, Perelman, Lott, Sturm 1, Sturm 2, Case}. In 2006, Chang and co-workers have shown how this space may welcome conformal (Weyl) transformations, and can help one construct conformally invariant geometrical objects~\cite{Chang}. 

Recently, inspired by the work of Chang et al., we have proposed a theory of gravitation consistent with conformal symmetry in the context of metric measure space~\cite{N1}. Indeed, by replacing Riemann space with metric measure space, a generalization of the Einstein equation is obtained, wherein the density function $f$ appears as a conformal degree of freedom. An interesting feature of this space is that, despite having conformal structure, metricity condition and integrability are retained. In another work~\cite{Graf}, Graf has used metric measure space (volume manifold) to formulate a geometric dilaton gravity. There, density function $f$ is considered as a dilation field and it is shown that some of the singularity problems can be avoided in this framework.

This manuscript is organized as follows. In Sec.~2, following~\cite{N1}, we present a conformally invariant theory of gravitation. This includes discussions about the field equations, the identities, the geodesic equation and the weak equivalence principle. In Sec.~3, some of the ideas about conformal transformations in gravitational theories, gauge fixing and symmetry breaking are discussed. We briefly review the de Broglie-Bohm interpretation of quantum mechanics in Sec.~4, and in the light of that, a behavior for the density function $f$ in quantum scale is proposed. In Sec.~5, it is shown how metric measure space may be regarded as an alternative to the integrable Weyl geometry. The manuscript closes in Sec.~6 with a brief summary.

\section{Conformally Invariant Theory of Gravitation in Metric Measure Space}

\subsection{Basic definitions}

Metric measure space is a triple $(M^n, g, dm)$ where $(M,g)$ is a (Pseudo-) Riemannian manifold and $dm$ is a smooth measure on $M$. The measure $dm$ is defined as $e^{-f} dvol(g)$ with $f$ being a density function. A tensor field $T$ is a conformally invariant tensor of weight $s$, if under conformal transformation $\hat{g}=e^{2w}g$, it transforms as $\hat{T}=e^{sw}T$. If one considers $s^\prime$ as the conformal weight of measure $dm$, it is easy to show that under $\hat{g}=e^{2w}g$, the density function $f$ satisfies the relation $\hat{f}=f+(n-s^\prime)w$. As a result, a conformally invariant geometrical object $\mathcal{I}_s(g,f)$ of conformal weight $s$, is defined as
\begin{equation}\label{a0}
\mathcal{I}_s(g,f)= \exp{(\frac{fs}{n-s^\prime})} I(\exp{(\frac{-2f}{n-s^\prime})}g),
\end{equation}
where $I(g)$ is its Riemannian invariant counterpart. Therefore, in metric measure space, the Ricci tensor of conformal weight $s_1$ is 
\begin{eqnarray}
\mathcal{R}_{\mu\nu}& =& \exp{(\frac{fs_1}{n-s^\prime})}[ R_{\mu\nu} + \frac{n-2}{n-s^\prime}\nabla_\mu \nabla_\nu f +\frac{n-2}{(n-s^\prime)(n-s^\prime)}\nabla_\mu f \nabla_\nu f \nonumber \\
&+& \frac{1}{n-s^\prime}\nabla^\alpha \nabla_\alpha f g_{\mu\nu} -\frac{n-2}{(n-s^\prime)(n-s^\prime)}\nabla^\alpha f \nabla_\alpha f g_{\mu\nu}],
\end{eqnarray}
and the Ricci scalar $\mathcal{R}$ of conformal weight $s_2$ is
\begin{eqnarray}\label{2}
\mathcal{R} = \exp{(\frac{s_2 f+2f}{n-s^\prime})}[R + 2(\frac{n-1}{n-s^\prime})\nabla^\mu \nabla_\mu f -\frac{(n-1)(n-2)}{(n-s^\prime)(n-s^\prime)}\nabla^\mu f \nabla_\mu f]. 
\end{eqnarray}

In metric measure space, there is a hint about the operators which have the integral based definitions~\cite{Chang2}. Indeed, since in this space, the measure is not the canonical volume element $dvol(g)$, these operators have different form from those of Riemann space. For instance, the divergence operator acting on a symmetric (0,2) tensor $X$ is defined as
\begin{equation}
\int \langle X, \nabla \vartheta \rangle dm = -\int \langle div X, \vartheta \rangle dm,
\end{equation}
where $\vartheta$ is a (1,0) tensor. Integrating by parts, which involves the measure $dm$, leads to the relation 
\begin{equation}\label{7n}
div X= \nabla_{\mu} X^{\mu\nu}-\nabla_\mu f X^{\mu\nu},
\end{equation}
for the divergence of a tensor $X$. In another definition, divergence is defined as the Lie derivative of an n-form. In Riemann space, the n-form is selected as volume form, while in metric measure space, the corresponding choice is the measure $dm$. In~\cite{N1}, it has been discussed, in detail, how this definition leads to the same relation as (\ref{7n}). 

\subsection{Dynamical equation and identities}

The conformally invariant action for gravitation in metric measure space is proposed as
\begin{equation} \label{b}
S= \int [\frac{1}{2\kappa} (\mathcal{R}- 2\exp{(-f)}\Lambda)+\mathcal{L}_{matter}(g,f, \psi)] dm,
\end{equation}
where $\kappa$ and $\Lambda$ are constants and $\psi$ is the matter field~\cite{N1}. There, it has been clarified that for the results to be in accordance with the integrable Weyl geometry, one should set conformal weight of $\mathcal{R}$ as -2, i.e. $s_2=-2$. In addition, in order for the action (\ref{b}) to be conformally invariant of weight zero, the conformal weight of $dm$ is set to $s^{\prime}=-s_2$. Hence, in this theory, $\mathcal{R}_{\mu\nu}$ and $\mathcal{R}$ have the weights $s_1=0$ and $s_2=-2$, respectively.

By variation of the action with respect to metric $g$, the equation
\begin{equation}\label{c}
\mathcal{R}_{\mu\nu}-\frac{1}{2} \mathcal{R} g_{\mu\nu}=-\exp{(-f)}\Lambda g_{\mu\nu}+\kappa \mathcal{T}_{\mu\nu}
\end{equation}
is obtained. The variation of (\ref{b}) with respect to the density function $f$ yields
\begin{equation}\label{d}
\frac{-1}{2\kappa}(\mathcal{R})+\frac{1}{2\kappa}(4\exp{(-f)}\Lambda)+\frac{1}{\exp{(-f)}}\frac{\delta(\exp{(-f)}\mathcal{L}_{matter})}{\delta f}=0,
\end{equation}
as well. The action has the conformal and diffeomorphism symmetries and each symmetry leads to an identity. The identity originated from conformal invariance of the action is derived as
\begin{equation}\label{19}
\mathcal{T}=\frac{-2}{\exp{(-f)}}\frac{\delta(\exp{(-f)}\mathcal{L}_{matter})}{\delta f},
\end{equation}
which shows that equation (\ref{d}) is the trace of equation (\ref{c}). In Riemann space, the identity resulted from diffeomorphism invariance of the gravitational part of the action (\ref{b}), is called contracted second Bianchi identity. Here, this symmetry leads to a generalized Bianchi type identity
 \begin{equation}\label{16}
\nabla_{\mu} \mathcal{R}^{\mu\nu}-\nabla_{\mu}f \mathcal{R}^{\mu\nu} -\frac{1}{2}\nabla^{\nu} \mathcal{R} =0.
\end{equation}
On the other hand, the diffeomorphism invariance of the matter part of the action gives
\begin{equation}\label{16sd}
\nabla_{\mu}\mathcal{T}^{\mu\nu}-\nabla_\mu f\mathcal{T}^{\mu\nu}+\frac{1}{2}\nabla^\nu f \mathcal{T}=0,
\end{equation}
which can be considered as a generalization of the conservation of energy-momentum tensor. To derive the two last relations, recall that the divergence operator in metric measure space is not simply achieved by a contraction of covariant derivative~\cite{N1}. In the next subsection, it is shown that this identity can be used to construct a conformally invariant geodesic equation.

It is notable that, we did not follow the assumption of Chang et al.~\cite{Chang} which has considered a fixed measure $dm$ under diffeomorphism. The reason is that the generalized field equation (\ref{c}) should reduce to the Einstein equation in the condition of $f=const$. Consequently, in this theory the equivalence conformal class $[(g,f=c)]$, for $c$ being a constant, resembles general relativity (GR). In other words, GR is retrieved by all the pairs $(\hat{g}=e^{2w}g, \hat{f}=c+(n-2)w)$, where the conformal factor $w$ is a constant. Indeed, here, the conformal transformations are reduced to a global change of units on the manifold. 

\subsection{Geodesics}

So far, we have constructed conformally invariant gravitational equations on a manifold with metric $g$ and density function $f$. To have a conformally invariant theory of gravitation, it is natural to demand geodesics which respect conformal invariance. One can define two kinds of geodesics on manifold $M$, namely metric geodesics and affine geodesics which are shortest and straightest possible lines, respectively. The definition of metric geodesics, as the name indicates is related to the metric, while the affine geodesics are constructed by connections. However, in the context of GR, geodesic equation can be derived from the diffeomorphism symmetry of the action, without any additional requirement. Here, following almost a similar procedure as in GR, the conformally invariant geodesic equation in the context of metric measure space is obtained. Moreover, it will be shown that in terms of conformal invariance, only the metric geodesic can be constructed compatible with this geodesic. 

\subsubsection{Deriving geodesics via generalized conservation of energy-momentum tensor} 

The diffeomorphism invariance of matter part of the action leads to the generalized conservation of energy-momentum tensor (\ref{16sd}). In what follows, we use this relation to derive the geodesic equation in metric measure space. Let $\mathcal{T}^{\mu\nu}$ be the energy-momentum tensor for an ideal fluid with pressure $p=0$, that is 
\begin{equation}\label{p}
\mathcal{T}^{\mu\nu}=h(f) \rho \overline{u}^\mu \overline{u}^\nu. 
\end{equation}
In this relation, $\overline{u}^\mu=\frac{dx^\mu}{d\lambda}$ where $d\lambda=e^\frac{f}{2} d\tau$, $\rho$ is the energy density and $h(f)$ is a function of density function $f$. As $\mathcal{T}_{\mu\nu}$ satisfies the relation (\ref{16sd}), one has
\begin{equation}\label{q}
\nabla_{\mu}(h(f) \rho \overline{u}^\mu \overline{u}^\nu)-\nabla_\mu f (h(f) \rho \overline{u}^\mu \overline{u}^\nu)-\frac{1}{2}\nabla^\nu f (h(f) \rho c^2 \exp{(-f)}) =0.
\end{equation}
On the other hand, conservation of the quantity of the fluid requires
\begin{equation}
div(h(f) \rho \overline{u})=0,
\end{equation}
which in metric measure space is read as
\begin{equation}\label{r32}
\nabla_\mu(h(f) \rho \overline{u}^\mu)-\nabla_\mu f (h(f) \rho \overline{u}^\mu)=0. 
\end{equation}
Multiplying (\ref{r32}) by $\overline{u}^\nu$ and considering (\ref{q}), one gets 
\begin{equation}\label{t}
\overline{u}^\mu \nabla_\mu \overline{u}^\nu-\frac{1}{2}c^2\exp{(-f)} \nabla^\nu f=0.
\end{equation}
The above relation in the language of coordinate becomes
\begin{equation}\label{s1}
\frac{d^2 x^\mu}{d\lambda^2}+\Gamma^\mu_{\nu\rho} \frac{dx^\nu}{d\lambda}\frac{dx^\rho}{d\lambda}-\frac{1}{2}c^2\exp{(-f)}\nabla^\mu f=0,
\end{equation}
where $\Gamma^\mu_{\nu\rho}$ denotes the Levi-Civita connection. It is interesting to note that the derived geodesic are invariant under transformations $\hat{g}=e^{2w}g$ and $\hat{f}=f+2w$. 

\subsubsection{Metric geodesics}

Relation (\ref{a0}) suggests $d\mathcal{S}=e^{\frac{-f}{2}} ds$ as the conformally invariant space-time separation of weight zero in metric measure space. Therefore, one expects that the conformally invariant geodesic equation achieved by optimizing the integral  
\begin{equation}
\mathcal{I}= \int_p^q \exp{(\frac{-f}{2})} ds,
\end{equation}
for fixed endpoints $p$ and $q$. After some mathematical manipulations, one ends up with
\begin{equation}\label{k}
\frac{d^2 x^\mu}{d\tau^2}+\Gamma^\mu_{\nu\rho} \frac{dx^\nu}{d\tau}\frac{dx^\rho}{d\tau}-\frac{1}{2}\nabla_\nu f \frac{dx^\nu}{d\tau}\frac{dx^\mu}{d\tau}-\frac{1}{2}c^2 \nabla^\mu f=0. 
\end{equation}
The third term can be removed by an appropriate reparametrization, namely $d\lambda=e^\frac{f}{2} d\tau$. The geodesic equation in the new parametrization then becomes
 \begin{equation}\label{l}
\frac{d^2 x^\mu}{d\lambda^2}+\Gamma^\mu_{\nu\rho} \frac{dx^\nu}{d\lambda}\frac{dx^\rho}{d\lambda}-\frac{1}{2}c^2\exp{(-f)} \nabla^\mu f=0,
\end{equation}
which is the same as the conformally invariant geodesic equation (\ref{s1}). 

\subsubsection{Affine geodesics}

As mentioned before, an affine geodesic is defined as the straightest possible line; indeed it is a privileged curve along which the tangent vector is parallelly transported. If the manifold is endowed with a metric, there exists a preferred connection according to which a vector is parallely transported. In the case of Riemannian geometry, this preferred connection is the Levi-Civita connection and is characterized by being 1) metric compatible and 2) torsion free. In metric measure space, similar to the Riemannian geometry, the Levi-Civita connection $\Gamma^\mu_{\nu\rho}$ is the preferred connection, since the two conditions above are held. Therefore, the affine geodesic equation in this space is
\begin{equation}\label{e1}
\frac{d^2 x^\mu}{d\tau^2}+\Gamma^\mu_{\nu\rho} \frac{dx^\nu}{d\tau}\frac{dx^\rho}{d\tau}=0,
\end{equation}
which is not conformally invariant.
 
\subsection{Weak equivalence principle}

Free fall of a test particle in metric measure space follows the conformally invariant geodesic equation (\ref{l}). It indicates that the motion of a test particle in a gravitational field is independent of its mass and composition. Hence, the weak equivalence principle in this space is held. However, as a result of the term $\frac{1}{2}c^2e^{-f} \nabla^\mu f$, the particles do not follow the Riemannian geodesics. 

\section{Conformal Freedom in Gravitational Theories}

The footprint of conformal transformation can be seen in different fields of physics, such as gravitational theories, conformal field theory (CFT), string theory and statistical mechanics. It is useful to clarify that there are two different transformations in the literature which both are called ``conformal transformation''. One of them, is the conformal (Weyl) transformation and is a local rescaling of the metric. This transformation usually is used in gravitational theories. The other, is the conformal coordinate transformation which is in fact a diffeomorphism​ and is the symmetry of conformal field theory. As important aspects of the latter, one can recall the AdS/CFT correspondence~\cite{Ads} and the Polyakov formalism in string theory~\cite{string}. Although conformal (Weyl) transformation and conformal coordinate transformation have essentially different structures, there are theorems which illustrate an interesting connection between them~\cite{Zumino, Blagojevic, Faci}. For instance, it is stated that if an action in Riemann space, is invariant under Weyl transformations, then when restricted to Minkowski space, it is invariant under conformal coordinate transformations. 

Conformal (Weyl) transformations typically appear with two different roles in classical theories of gravitation, i.e. either as a mathematical tool or as a symmetry transformation. In the following, we will briefly mention these two roles.  

\subsection{Conformal transformation as a mathematical tool} 
 
Conformal transformation can be used as a mathematical tool to map complicated equations of motion into mathematically equivalent sets of equations that can be solved more easily. These transformations, for instance, appear in scalar-tensor theories of gravitation which are of the most popular alternatives to GR. Although the scalar-tensor theories are not in general conformally invariant, they have some functional degrees of freedom which allow one to write down the equations, by appropriate redefinition of fields, in different conformal frames. Among these conformal frames, the Einstein and the Jordan frames are discussed more frequently. There exist wide variety of articles in the literature which discuss about either the equivalence or nonequivalence of the scalar-tensor theories of gravitation formulated in different conformal frames. It seems that the viewpoint which assert that these frames are equivalent, is more consistent. However, one should note that these equivalence is held as long as one does not treat any convention arising in the interpretation of the theory as fixed. For instance, if one assumes the metric in both the Einstein and the Jordan frames to be a metric which is measured by the atomic clocks, then these frames will not be physically equivalent anymore~\cite{flanagan, faraoni}. Furthermore, in the literature, one also finds the idea that the conformal equivalence is held just at the classical level, and quantum considerations can make difference between different conformal frames~\cite{flanagan, faraoni2}. 

\subsection{Conformal transformation as a symmetry transformation}

Interest in conformal transformations as symmetry transformations arose after the Weyl's paper~\cite{Weyl 2}, which was the early form of the gauge theories. Weyl, in addition to imposing conformal symmetry on a gravitational theory, tried to build a geometrical theory of electromagnetism. Although Weyl's theory was physically indefensible due to the problem of non-integrability, the beauty of the theory encouraged many authors to modify and even reinterpret it. For instance, Dirac tried to revive Weyl's theory by using a new scalar field $\beta$. Indeed, he resolved the problem of non-integrability by introducing the notions of atomic and gravitational distances~\cite{Dirac}. As other well-known conformally invariant theories of gravitation in the context of (integrable) Weyl geometry, one can mention the works of Rosen~\cite{Rosen} and Canuto et al.~\cite{Ca1,Ca3}. 

On the other hand, there is another way of achieving conformal symmetry in gravitational theories. In this approach, a non-minimal coupling of a Brans-Dicke-like scalar field $\phi$ to the Ricci scalar is considered in the way that the action be conformally invariant (up to a total derivative). For instance, Bars, Steinhardt and Turok by using this method, have provided conformally invariant version of the standard model coupled to gravitation~\cite{Bars}. Indeed, they have presented a theory in which gravitation coupled to fermions, Higgs and other gauge bosons, while maintaining conformal symmetry. The geodesically complete cosmological solutions have been also discussed as an important application of this conformally invariant theory. In another recent work, Scholz has investigated the probable link between the Brans-Dicke-like scalar field $\phi$ and the Higgs field~\cite{Scholz2}. His work is explicitly formulated in the context of Weyl geometry, contrary to the approach of~\cite{Bars}. 

After Dicke's paper~\cite{Dicke}, conformal transformations are usually interpreted as coordinate-dependent unit transformations. This point of view may lead to the idea that conformal symmetry just reflects the freedom of local choice of units. In other words, conformal symmetry indicates the fact that an observer is not obliged to use a special and fixed system of units which is rigidly attached to spacetime. It is traditional to suppose mass and constants in the theory to be changed under conformal transformations. Sometimes, the conformal weights of these physical constants are chosen based on their physical units. For example, conformal weights of the gravitational constant $G$ and mass are considered as 2 and -1, respectively. However, it is not a general trend, see for instance~\cite{Ca1}. 

So far, we have mentioned some of the gravitational theories underlying conformal symmetry. Generally, the authors, in dealing with this symmetry, follow two roughly related methods, which are gauge fixing and symmetry breaking. They are related since the result of the both approaches are the same: conformal symmetry is not held anymore.  

\subsubsection{Gauge fixing}

In this viewpoint, a gauge condition is applied to eliminate the arbitrariness of a conformally invariant theory. For instance, Canuto and co-workers imposed a gauge condition by using Dirac's Large Number Hypothesis (LNH) in their theory, so that the gravitational phenomena can be described in atomic units~\cite{Ca1}. Indeed, they have believed that GR has been formulated in gravitational units and, in order to describe gravitational phenomena in atomic units, one should determine the relationship between these system of units. Therefore, they stated that LNH which deals with the ratio of constants in these units, seems to be a proper gauge condition in cosmology. There are other efforts to fix the conformal gauge, mostly via cosmological observations~\cite{Ca2, Maeder, Scholz}. In summary, in the gauge fixing method, an appropriate gauge condition for each situation can be applied without any specific mechanism, just regarding different paradigms such as establishing hypotheses or considering observations. 

\subsubsection{Symmetry breaking} 

This viewpoint is asserting that the conformal symmetry is a kind of symmetry that is broken, see for instances~\cite{Deser, Zee, Zee2, Ohanian, Kleinert}. In this setting, a symmetry breaking mechanism is needed and a standard gravitational theory is constructed by symmetry breaking of {\it a priori} conformally invariant theory. Usually, two approaches occurring in the literature supporting this idea. In the first approach, conformal symmetry breaking happens, exactly when the zero-mass particles acquire mass. In the second approach, however, a separate mechanism for this symmetry breaking, independent of the electroweak gauge symmetry breaking, is assumed.

\section{Gauge fixing in quantum scale}

The presented theory in this manuscript respects conformal invariance in the context of metric measure space. Although this framework has features which are not usual, one may show that some results of the proposed theory, by a redefinition of fields, match with other theories. For instance in~\cite{N1}, we have demonstrated how the Dirac field $\beta$, in Canuto et al.'s theory~\cite{Ca1}, can be equivalent to the term $e^{-f}$ in our theory formulated in metric measure space. The constants in both theories are included in the Lagrangian of matter. The fact that this Lagrangian is considered as a function of field $f$, leads the trace of the energy momentum tensor not to be zero necessarily. However, here, we will not follow the approach of~\cite{Ca1} in determining the conformal degree of freedom. Instead of gauge fixing via cosmological observations or LNH for instance, we propose a physical selection for the density function $f$ via quantum observations. Indeed, we choose a particular gauge of the broken symmetry, valid in quantum scale. Our motivation for gauge fixing in quantum scale is that the quantum mechanics is a well-established theory, and its predictions are in good agreement with experiments. To do so, we choose the Bohm's interpretation of quantum mechanics since it deals with geometrical trajectories of particles, while providing a physically equivalent theory to quantum mechanics, as well. 

In what follows, the Bohm's quantum mechanics and one of the relativistic approaches of it~\cite{shojai1}, are reviewed. Afterwards, we investigate limits under which the geodesic equation (\ref{s1}) coincides with the Bohmian trajectories. That is, by comparing the trajectories in Bohm's interpretation of quantum mechanics with the geodesics in metric measure space, we find a relation for density function $f$ in quantum scale. It is shown that not only in non-relativistic regime, but also in the relativistic regime the geodesics of the Bohm's theory of quantum and metric measure space have similar structures.

\subsection{de Broglie-Bohm interpretation of quantum mechanics}  

The de Broglie-Bohm theory mainly rely on a causal interpretation of quantum mechanics. In this approach, a physical system comprises a propagating wave together with a point particle which moves under the guidance of the wave. The wave function of a particle is written as $\psi=\sqrt{\rho} \exp[{\frac{iS}{\hbar}}]$, where $S$ is Hamilton-Jacobi function and $\rho$ is the ensemble density of the system. For a detailed discussion of Bohmian quantum mechanics one can see~\cite{Bohm, Holand, Li, Caroll}.

In the de Broglie-Bohm interpretation of quantum mechanics, each particle has a definite trajectory, determined by 
\begin{equation}\label{qq}
\frac{d\overrightarrow{x}}{dt}=\frac{1}{m} \overrightarrow{\nabla} S(x,t),
\end{equation}
where the background is the 3D euclidean space. On the other hand, by substituting the wave function $\psi=\sqrt{\rho} \exp[{\frac{iS}{\hbar}}]$ into Schrödinger's equation and separating imaginary and real parts, the relations 
\begin{equation}\label{r}
\frac{\partial \rho}{\partial t}+\nabla . (\frac{\overrightarrow{\nabla} S}{m}\rho)=0
\end{equation}
and
\begin{equation}\label{qb}
\frac{\partial S}{\partial t}+ \frac{|\overrightarrow{\nabla} S^2|}{2m}+V-\frac{\hbar^2}{2m}\frac{\nabla^2 \sqrt{\rho}}{\sqrt{\rho} }=0,
\end{equation}
are obtained. The equation (\ref{r}) is continuity equation for the probability density $\rho$, and the equation (\ref{qb}) is named the quantum Hamilton-Jacobi equation, since it is equal to the classical Hamilton-Jacobi equation with the additional quantum potential
\begin{equation}\label{s}
Q=-\frac{\hbar^2}{2m }\frac{\nabla^2 \sqrt{\rho}}{\sqrt{\rho}.}
\end{equation}
By applying the gradient operator on both sides of relation (\ref{qb}) and considering the relation (\ref{qq}), one gets 
\begin{equation}\label{t}
m\frac{d^2\overrightarrow{x}}{dt^2}=-\overrightarrow{\nabla}(V+Q),
\end{equation}
as the equation of motion of a particle in non-relativistic quantum mechanics. This relation is Newton's second law which shows that in addition to the classical force $-\overrightarrow{\nabla} V$, a quantum force $-\overrightarrow{\nabla} Q$ is applied to the particle.

Bohm and de-Broglie constructed a relativistic version of this theory by setting $\psi=\sqrt{\rho} \exp[{\frac{iS}{\hbar}}]$ in the Klein-Gordon equation. However, their theory suffers from difficulties related to the definition of a field, named quantum mass $\mathcal{M}$. In~\cite{shojai1}, a modified definition of $\mathcal{M}$ has been presented which resolves the problem. The authors in ~\cite{shojai1}, have developed a Bohm's approach to the Klein-Gordon equation wherein the equation of motion has the correct non-relativistic limit. In their approach, the quantum Hamilton-Jacobi equation for a relativistic spin-less particle is given by
\begin{equation}
g^{\mu\nu}\nabla_\mu S \nabla_\nu S= \mathcal{M}^2 c^2,
\end{equation}
in which the quantum mass $\mathcal{M}$ is defined as
\begin{equation}
\mathcal{M}^2=m^2 \exp{(\mathcal{Q})},
\end{equation}
with the relativistic quantum potential $\mathcal{Q}$ being
\begin{equation}\label{lol}
\mathcal{Q}=\frac{\hbar^2}{m^2 c^2}\frac{(\frac{1}{c^2}\frac{\partial^2}{\partial t^2}-\nabla^2) \sqrt{\rho} }{\sqrt{\rho} }.
\end{equation}
In this setting, the relation
\begin{equation}\label{lo}
\mathcal{M} \frac{d^2x^\mu}{d\tau^2}+\mathcal{M} \Gamma^\mu_{\nu\rho} \frac{dx^\nu}{d\tau}\frac{dx^\rho}{d\tau}=\frac{1}{2}(c^2 g^{\mu\nu}-\frac{dx^\mu}{d\tau}\frac{dx^\nu}{d\tau})\nabla_\nu \mathcal{M}, 
\end{equation}
determines the motion of a particle, which is indeed the geodesic equation in Riemann space, influenced by a quantum force~\cite{Li, Caroll, shojai1}. 


\subsection{Trajectory of a microscopic particle in weak field limits}

Consider a stationary weak gravitational field and a test particle which moves slowly in comparison with the speed of light. Slowly moving condition $\varepsilon \equiv \frac{v}{c} \ll1$, can be written as
\begin{equation}
\frac{dx^i}{d\tau}= \varepsilon \frac{dx^0}{d\tau},
\end{equation}
where the index $i$ runs from 1 to 3. This implies that $\frac{\varepsilon}{\delta x^i}\sim \frac{1}{\delta x^0}$. Therefore for any function $f$, one has 
\begin{equation}\label{th}
\varepsilon \frac{\partial f}{\partial x^i}\sim  \frac{\partial f}{\partial x^0}. 
 \end{equation}
On the other hand, for weak fields, we may adopt a nearly Cartesian coordinate system wherein 
\begin{equation}\label{t1}
g_{\mu\nu}=\eta_{\mu\nu}+ h_{\mu\nu},
\end{equation}
with $h_{\mu\nu} \propto O(\varepsilon)$. Let us assume $\nabla_i f \propto O (\varepsilon)$ in quantum scale. Then by relation (\ref{th}), one finds $\nabla_0 f \propto O (\varepsilon^2)$. Putting it all together, to the leading order of $\varepsilon$, the geodesic equation (\ref{s1}) becomes 
\begin{equation}\label{u}
\frac{d^2 x^\mu}{d\tau^2}=-\Gamma^\mu_{00}\frac{dx^0}{d\tau}\frac{dx^0}{d\tau}+\frac{1}{2}c^2 \eta^{\mu i} \nabla_{i} f.
\end{equation}
The component $\mu=0$ of the above relation is 
\begin{equation}
\frac{d^2 x^0}{d\tau^2}=0,
\end{equation}
which implies $t=\tau$ as a candidate. Considering stationary field condition $\partial_0 g_{\mu\nu}=0$, the components $\mu=i$ to the leading order, can be written as 
\begin{equation}\label{w}
\frac{d^2 x^i}{dt^2}=\frac{c^2}{2}\partial^i(h_{00}+ f). 
\end{equation}
By multiplying both sides of above relation by $m$ and setting $V=-\frac{mc^2}{2} h_{00}$ and $Q=-\frac{mc^2}{2}f$, one has
\begin{equation}\label{p1}
m\frac{d^2 x^i}{dt^2}=-\partial^i(V +Q).
\end{equation}
As in GR, the term $-\partial^i V$ can be considered as the Newtonian gravitational force, while the term $-\partial^i Q$ is the additional force induced by the density function $f$. 

Relation (\ref{p1}) shows similar structure with the trajectory of a particle in non-relativistic Bohmian quantum mechanics, namely relation (\ref{t}). However, one should note that the potential $Q$ in the above relation is resulted from the geometrical density function $f$, while in the relation (\ref{t}) it is a quantum potential. Nevertheless, we can deduce that the geodesic equation in metric measure space, in appropriate limits, can reproduce the Bohmian trajectories. These limits include Newtonian approximation as well as $\nabla_i f \propto O (\varepsilon)$, which the latter can be dubbed as quantum scale condition. In addition, the relation $\nabla_i f \propto O (\varepsilon^2)$ should be satisfied in classical limit, to be held in agreement with the the classical limit condition of Bohm's theory, namely the relation $\overrightarrow{\nabla} Q \ll \overrightarrow{\nabla} V$.  
 
It is interesting to compare the path taken by a particle in metric measure space, relation (\ref{k}), with its correspondence in the relativistic regime of Bohm's theory, relation (\ref{lo}). However, it should be noted that the equations in this manuscript have been adapted to the metric signature $(- + + +)$ which is opposite to what it has been used in~\cite{shojai1}. Hence, before comparing the mentioned relations, a minus sign for the term $c^2 g^{\mu\nu}$ in (\ref{lo}) should be considered. Consequently, one finds that the term $e^{-f}$ can be considered as the quantum mass $\mathcal{M}$, up to a constant coefficient. Setting this coefficient to $m$ leads to $\mathcal{Q}=-f$, which is in accordance with the relation $Q=-\frac{mc^2f }{2}$, obtained for non-relativistic limit. This accordance does not come as a surprise, since the authors of~\cite{shojai1} have constructed the covariant equation of motion (\ref{lo}) in such a way that the correct non-relativistic limit (\ref{t}) emerges. 

In this section, we have followed the Bohm's approach for non relativistic quantum mechanics and the work of A. Shojai and F.Shojai~\cite{shojai1} for relativistic generalization of it. However, it should be noted that there are some other geometrical approaches for having a physically equivalent theory to quantum mechanics. For instance, in~\cite{santa0}, Santamato has obtained trajectories of particles by classical mechanics, while initial positions are assumed to be random. Recently, by extending that work, Santamato and De Martini have derived the Dirac's equation for a particle with spin~\cite{santa1} and investigated the spin statistic problem~\cite{santa2}. In this Bohm-type quantum mechanics, the origin of quantum effects is a feedback between geometry and dynamics and hence the quantum mechanical force is related to the Weyl connection. In comparison, in our work, Bohm interpretation of quantum mechanics is used to apply gauge condition in metric measure space. Consequently, the quantum potential is obtained proportional to the density function.

As another example of geometrical approach, Novello, Salim and Falciano~\cite{salim} have derived relations of non-relativistic Bohmian quantum mechanics by modifying Euclidean space. In their work, it is shown that by applying Palatini-like procedure, the 3D Weyl integrable space (Q-wis) appears more natural than the ad hoc 3D Euclidean space, in the micro-world scale. In Weyl space, even if the metric is considered Euclidean, the space is still curved. Hence, with the assumption of Euclidean background, it is shown that the non-relativistic Bohmian quantum potential can be identified with the Weyl curvature scalar in their framework. However, in this manuscript, the starting point is different from~\cite{salim} and in addition, due to gravitational concerns, we have considered the metric to be dynamical. 
\section{Metric Measure Space as an Alternative to Integrable Weyl Space}

In 1918, Hermann Weyl presented a generalization of (Pseudo-) Riemannian geometry which is known as Weyl geometry~\cite{Weyl1}. Weyl space $(M^n, g,\kappa)$ is an n-dimensional smooth manifold $M$ endowed with a metric $g$ and a 1-form $\kappa$, which is called length connection. In this space, quantities measured at infinitesimally close points can be metrically compared indirectly, only after using $\kappa$ to transport length unit along a path connecting these points. Weyl realized that this structure is achieved if one assumes the non-metricity condition
\begin{equation}\label{f}
\overline{\nabla}_\lambda g_{\mu\nu}= -2 \alpha \kappa_\lambda g _{\mu\nu},
\end{equation}
where $\alpha$ is an arbitrary constant. Hence, Weyl geometry has two essential features, namely non-metricity and non-integrability. Relation (\ref{f}) is invariant under the joint transformations $\hat{g}_{\mu\nu}=e^{2w}g_{\mu\nu}$ and $ \hat{\kappa}_\mu=\kappa_\mu-\frac{1}{\alpha}\nabla_\mu w$. Consequently, the geometrical tensors constructed by the generalized covariant derivative $\overline{\nabla}$ are also invariant under these joint transformations. Indeed, conformally invariant geometrical objects in this framework are obtained by substituting the covariant derivative $\overline{\nabla}$ in the Riemannian definitions. 

Integrable Weyl space is a subclass of Weyl space wherein $\kappa$ is considered as the gradient of a scalar field, say $\varphi$. Therefore there is no change in the length of a vector along a closed path, and this makes integrable Weyl space more consistent to formulate a physical theory than the genuine Weyl geometry. The generalized conformally invariant Ricci tensor and scalar in this space are
\begin{equation}\label{23fd}
\overline{R}_{\mu\nu}= R_{\mu\nu} - (n-2)\alpha \nabla_\mu \kappa_\nu +(n-2)\alpha^2 \kappa_\mu  \kappa_\nu  - \alpha \nabla^\sigma \kappa_\sigma  g_{\mu\nu} -(n-2)\alpha^2 \kappa^\sigma \kappa_\sigma  g_{\mu\nu}
\end{equation}
and 
\begin{equation}\label{23s}
\overline{R}= R-2(n-1)\alpha \nabla_\mu \kappa^\mu-(n-2)(n-1) \alpha^2 \kappa _\mu \kappa^\mu,
\end{equation}
respectively. Affine and metric geodesics in integrable Weyl space are also defined. Let's denote the connection coefficients of the covariant derivative $\overline{\nabla}$ by Weyl-Levi-Civita connection $\overline{\Gamma}^\sigma_{\mu\nu}$. If one parallel-transports a tangent vector $u=\frac{dx^\mu}{d\tau}$ by the connection $\overline{\Gamma}^\sigma_{\mu\nu}$ and takes into account the conformal weight of $u$, one ends up with 
\begin{equation}\label{i1}
\frac{d^2 x^\mu}{d\tau^2}+\Gamma^\mu_{\nu\rho} \frac{dx^\nu}{d\tau}\frac{dx^\rho}{d\tau}+\alpha \nabla_\nu \varphi  \frac{dx^\nu}{d\tau}\frac{dx^\mu}{d\tau}+\alpha c^2 \nabla^\mu \varphi =0
\end{equation}
as the conformally invariant affine geodesic. Moreover, the conformally invariant metric geodesic is obtained by variation of $I=\int e^{\alpha \varphi} ds$, with the final result being the same as (\ref{i1}). 

It is clear that integrable Weyl and metric measure spaces are constructed based on different assumptions. In integrable Weyl space, conformally invariant geometrical objects are constructed by Weyl-Levi-Civita connection $\overline{\Gamma}^\sigma_{\mu\nu}$, while metric measure space provides a framework to have conformally invariant geometrical objects without changing the preferred connection $\Gamma$. In spite of all differences, one can show that there is a duality between some of the relations in metric measure space and integrable Weyl space. Indeed, we have shown that by assuming  
\begin{equation}\label{mm}
\kappa_\mu=\frac{-1}{2\alpha}\nabla_\mu f,
\end{equation}
the joint transformations, the conformally invariant Ricci tensor and Ricci scalar, and the second Bianchi identity in these spaces are completely similar~\cite{N1}. The similarity of these relations, especially the second Bianchi identity, regarding the fact that they are obtained from completely different methods, is an interesting result. 

To complete our journey to compare metric measure space and integrable Weyl space, it is constructive to compare geodesics in these spaces. In Sec.~2, the conformally invariant geodesic equation (\ref{s1}) in the context of metric measure space is obtained. Comparing this relation with its counterpart in integrable Weyl space, relation (\ref{i1}), one finds that by assuming (\ref{mm}), these geodesics coincide, as well. However, it is worth noting that affine geodesic in metric measure space cannot naturally be made conformally invariant. This feature is in contrast to the integrable Weyl space where both metric and affine geodesics are conformally invariant. 


\section{Summary}

The aim of this paper is to complete what provided in~\cite{N1} to clarify all relevant aspects of metric measure space in formulation of a conformally invariant theory of gravitation. Here, we have investigated the conformally invariant geodesic equation and weak equivalence principle, in addition to the dynamical equations and the identities presented in~\cite{N1}. Moreover, we have demonstrated how a special behavior of density function $f$ in quantum scale can produce a similar effect on the trajectory of particles, as Bohm's quantum theory does. This outcome is in agreement with a belief which states Bohm's quantum mechanics may be understood on the basis of a geometric interpretation. In other words, it seems that Bohm's quantum theory can be regarded as the limiting regime of the conformal structure endowed in metric measure space. The suggested method in comparing of the geodesics in metric measure space with Bohmian trajectories, is indeed a physical selection of a conformal gauge in quantum scale. 
 

\end{document}